\documentclass[3p]{elsarticle}
\usepackage{lipsum}
\makeatletter
\def\ps@pprintTitle{%
 \let\@oddhead\@empty
 \let\@evenhead\@empty
 \def\@oddfoot{Accepted version, publisher version available at \url{https://doi.org/10.1016/j.bspc.2022.103538}\\}%
 \let\@evenfoot\@oddfoot}
\makeatother

\usepackage{lineno,hyperref}
\modulolinenumbers[5]

\journal{Biomedical Signal Processing and Control}

\usepackage{amsmath}
\usepackage{siunitx}
\usepackage{nicefrac}
\usepackage{caption}
\usepackage{subcaption}
\usepackage{algorithm} 
\usepackage{algpseudocode} 
\usepackage{color}
\definecolor{gray}{gray}{.5} 

\newlength{\inputIntent}
\usepackage{ulem}
\newcommand{\revision}[1]{#1}
\newcommand{\revdel}[1]{}
\newcommand{\revdelred}[1]{}

\begin{document}

\begin{frontmatter}
\title{Signal-to-noise ratio is more important than sampling rate in beat-to-beat interval estimation from optical sensors}



\author[mymainaddress]{Sebastian Zaunseder}
\author[anttiaddress]{Antti Vehkaoja}
\author[mymainaddress]{Vincent Fleischhauer}
\author[christophaddress]{Christoph~Hoog~Antink\corref{mycorrespondingauthor}}


\cortext[mycorrespondingauthor]{Corresponding author}
\ead{hoogantink@kismed.tu-darmstadt.de}

\address[mymainaddress]{Faculty of Information Technology, FH Dortmund, Sonnenstraße 96, 44139 Dortmund, Germany}
\address[anttiaddress]{Finnish Cardiovascular Research Center and Faculty of Medicine and Health Technology, Tampere University, Korkeakoulunkatu 3, 33720 Tampere, Finland}
\address[christophaddress]{KIS*MED (AI Systems in Medicine), TU Darmstadt, Merckstraße 25, 64283 Darmstadt, Germany}

\begin{abstract}
Photoplethysmographic Imaging (PPGI) allows the determination of pulse rate variability from sequential beat-to-beat intervals (BBI) and pulse wave velocity from spatially resolved recorded pulse waves. In either case, sufficient temporal accuracy is essential. 

The presented work investigates the temporal accuracy of BBI estimation from photoplethysmographic signals. Within comprehensive numerical simulation, we systematically assess the impact of sampling rate, signal-to-noise ratio (SNR), and beat-to-beat shape variations on the root mean square error (RMSE) between real and estimated BBI.

Our results show that at sampling rates beyond \SI{14}{\hertz} only small errors exist when interpolation is used. For example, the average RMSE is \SI{3}{ms} for a sampling rate of \SI{14}{\hertz} and an SNR of \SI{18}{dB}. Further increasing the sampling rate only results in marginal improvements, e.g. more than tripling the sampling rate to \SI{50}{\hertz} reduces the error by approx. \SI{14}{\%}. The most important finding relates to the SNR, which is shown to have a much stronger influence on the error than the sampling rate. For example, increasing the SNR from \SI{18}{dB} to \SI{24}{dB} at \SI{14}{\hertz} sampling rate reduced the error by almost \SI{50}{\%} to \SI{1.5}{ms}. Subtle beat-to-beat shape variations, moreover, increase the error decisively by up to \SI{800}{\%}.

Our results are highly relevant in three regards: first, they partially explain different results in the literature on minimum sampling rates. Second, they emphasize the importance to consider SNR and possibly shape variation in investigations on the minimal sampling rate. Third, they underline the importance of appropriate processing techniques to increase SNR. Importantly, though our motivation is PPGI, the presented work immediately applies to contact PPG and PPG in other settings such as wearables. To enable further investigations, we make the scripts used in modelling and simulation freely available. 
\end{abstract}

\begin{keyword}
Photoplethysmography, PPG, photoplethysmogram, Photoplethysmographic imaging, PPGI, sampling rate, sampling frequency, frame rate, signal-to-noise ratio, Monte Carlo simulation, imaging, wearable, pulse decomposition
\end{keyword}

\end{frontmatter}


\section{Introduction}
Photoplethysmographic Imaging (PPGI) has become a topic of active research in recent years. PPGI closely relates to conventional (contact) photoplethysmography (PPG) as it captures modulations of light absorption or reflection due to varying amount of blood~\cite{Allen2007}. Different from conventional PPG, which uses sensors attached to the skin, PPGI uses cameras to capture the superficial blood volume filling without any contact, thus enabling highly convenient monitoring~\cite{Antink2019,Zaunseder2018a}. Most applications of PPGI focus at pulse rate (PR) measurement~\cite{DeHaan2013,Poh2010,Woyczyk2021}, pulse rate variability (PRV) estimation~\cite{Iozzia2016,McDuff2015,Sun2013} or estimation of arterial oxygen saturation (SpO2)~\cite{Verkruysse2017,Moco2020}. Even beyond PR, PRV, and SpO2, PPGI bears relevant information. Recent works direct at pulse wave analysis from PPGI to reveal information regarding blood pressure and vascular state~\cite{Fleischhauer2020,Wang2020b}. Further, several research groups used PPGI to analyse pulse wave propagation (pulse transit time (PTT) or pulse wave velocity (PWV)), which also relates to blood pressure. Pulse wave propagation has been captured by a single camera considering spatially separated regions of interest (often face and palm)~\cite{Fan2020, Huang2017, Kaur2017,Jeong2016,Murakami2015,Saiko2021,Zaunseder2017} or by a combination of PPGI and other biosignal acquisition techniques~\cite{Kamshilin2016,Mamontov2020,Zhang2017d}.  

Its wide information content and convenient application conditions render PPGI a promising technology. However, PPGI also exhibits drawbacks. Even under ideal conditions, PPGI suffers from a low signal-to-noise ratio (SNR) compared to the conventional PPG. Subject movements or varying illumination conditions further decrease SNR. Moreover, PPGI typically has a limited temporal accuracy. Required minimum exposure time and huge amounts of data are responsible for frame rates typically being below 60 frames per second (fps), most often around \SI{25}{fps}~\cite{Antink2019,Zaunseder2018a}. While laboratory settings might allow higher frame rates for experimental studies~\cite{Jeong2016,Sun2013,Saiko2021,Trumpp2016}, real use cases as home monitoring would need to utilize inexpensive standard web cams at reduced frame rates.

Particularly with respect to PRV and analysis of pulse wave propagation, low frame rates pose a seemingly obvious problem. Both techniques rely on an exact estimation of beat-to-beat intervals (BBI). PRV requires the temporal difference between sequential beats whereas pulse propagation analysis requires the temporal difference between the same beat recorded at different measurement sites (the term \textit{beat-to-beat intervals} refers to both in the following description). While, to the best of our knowledge, there are no works investigating the effect of sampling rate on the temporal accuracy of obtained BBI with particular focus on PPGI, several authors have recently studied the impact of sampling rate on conventional PPG signals. The existing studies typically pursue a three-stage approach: (1) preprocessing, i.e. downsampling of original PPG data and sometimes upsampling/interpolation again (2) extraction of features from preprocessed data (BBI extraction and PRV parameters, occasionally morphological features), and (3) comparison of features from (2) against a reference (features from a PPG without downsampling or ECG derived features). Using such approach, Choi et al. proposed using at least \SI{25}{Hz} sampling rate~\cite{Choi2017a} for reliable PRV parameters. Beres et al. did find reliable results on PRV accuracy starting at sampling rates of \SI{20}{Hz}~\cite{Beres2021}. With respect to the accuracy of BBI and pulse waveform features, Fujita and Suzuki recommended using at least \SI{30}{Hz} for most parameters~\cite{Fujita2019}. Bent and Dunn as well as Pelaez-Coca et al. reported higher minimal sampling rates for PRV analysis: \SI{64}{Hz} in~\cite{Bent2021} and \SI{100}{Hz} in~\cite{Pelaez-Coca2021}, respectively, and thus at least seemingly contradicting results. However, according to Pelaez-Coca et al., using a fiducial point that combines multiple points from the pulse waveform and interpolation allows a reduction to \SI{50}{Hz} for reliable PRV parameters~\cite{Pelaez-Coca2021}. In line with this finding, Baek et al. showed that the impact of low sampling frequencies diminish if interpolation is applied to the sparsely sampled signals~\cite{Baek2017}. They found that a signal sampled at \SI{20}{Hz} almost yields the same results as the original signals at \SI{250}{Hz} when spline or parabolic interpolation is applied. Similarly, Beres et al. emphasize the positive effect of interpolation if a low sampling rates is used~\cite{Beres2021}. With respect to the absolute numbers given, it is very important to emphasize that the minimum sampling rate depends on the considered parameters. Some parameters have a high frequent signature, e.g. root mean square of the successive BBI, and require higher sampling rate than others like standard deviation of BBI. The given numbers reflect an attempt to provide the minimum sampling rate to assess parameters with an assumed high frequent signature from the respective papers, which typically provide different minimum sampling rates.

Though the investigations have common aim and structure, their findings differ. Multiple factors (apart from the use of interpolation or not) contribute to such differences. First of all, the criterion to decide on a sufficient sampling rate differs. Moreover, all aforementioned works conducted their studies using real world data. The extent of data was naturally limited (typically 60 subjects or less) and the considered populations, and thus temporal and morphological signal characteristics, show differences. Further, different recording equipment and non-uniform recording protocols were used. Thus, varying (and unspecified) signal qualities, amounts of noise and interference were present and likely have affected the results of each study. That being the case, none of the studies has systematically investigated the impact of such factors. In fact, the commonly employed approach, the usage of real data, which on the one hand is beneficial, has systematic limitations regarding investigations on the impact of noise.

As available results are ambiguous and the common approach has systematic limitations, this work complements existing works by comprehensive numerical simulations on the temporal accuracy of BBI estimation from PPG signals. Our analysis does not only consider the impact of sampling rate but, for the first time, systematically takes into account both, the effect of signal-to-noise ratio (SNR) and random variation in the pulse wave morphology. To that end, we combine a pulse de-/recomposition framework and Monte Carlo simulations into a versatile simulation environment. To the best of our knowledge, there are no comparable works: we are only aware of one study~\cite{Beres2019} that used simulated data. In this study, the PPG signals were modeled as frequency modulated cosine waves and thus highly simplified. Also, the study did not take varying SNR and beat-to-beat shape variations into consideration. Though PPGI and associated applications motivated our research, the presented considerations immediately apply to contact PPG as well. Particularly for the rapidly expanding field of PPG in consumer devices such as smart watches or optical heart rate monitors our investigation is highly relevant, as these devices typically show limited SNR and low sampling frequency to prolong battery life and optimize data management~\cite{Bent2021}. Regarding conventional PPG, to which direct most of the existing studies on minimum sampling rate, our analyses might help to explain reported differences as mentioned above and reveal aspects that received little attention so far. In order to enable more and deeper investigations in this regard, we release our data and all code to the public domain \url{https://github.com/KISMED-TUDa/PPG_Sim_SNR_Fs}. 

\section{Materials and Methods}

\subsection{Data\revision{ generation}}
The study uses simulated PPG pulse prototypes that are corrupted by varying levels of noise and that can undergo random shape variations. The following section details the employed procedure, which takes up ideas from~\cite{Fleischhauer2020}. 

\subsubsection{Prototype pulse generation}
\label{sec:prototype_generation}
\begin{figure}
    \centering
\begin{subfigure}[c]{4cm}
    \subcaption{Class 1}
    \includegraphics[width=\textwidth]{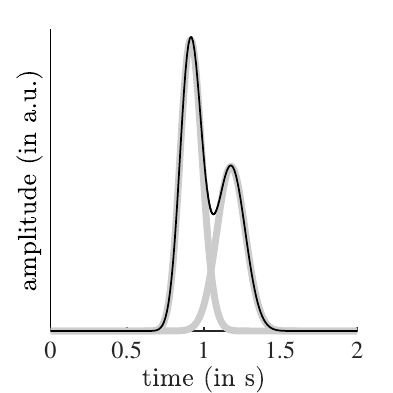}
    \newline
\end{subfigure}
\begin{subfigure}[c]{4cm}
    \subcaption{Class 2}
    \includegraphics[width=\textwidth]{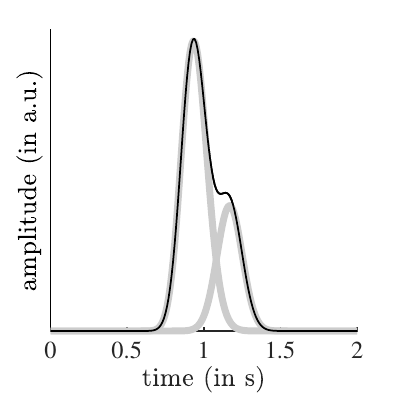}
    \newline
\end{subfigure}
\begin{subfigure}[c]{4cm}
    \subcaption{Class 3}
    \includegraphics[width=\textwidth]{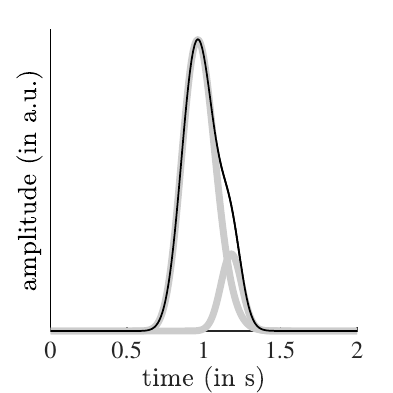}
\end{subfigure}
\begin{subfigure}[c]{4cm}
    \subcaption{Class 4}
    \includegraphics[width=\textwidth]{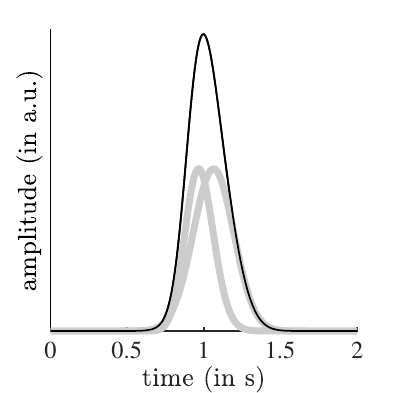}
\end{subfigure}
    \caption{Simulated prototype pulses according to the four Dawber classes~\cite{Dawber1973}. Pulses are generated by recomposition using two kernels (in gray) parametrized by a predefined parameter vector $\mathbf{\theta}$. Table~\ref{tab:PARAMETERS} provides the used parameter values.}
    \label{fig:CLEAN_PPG}
\end{figure}

We build up pulse prototypes by a recomposition of kernel functions. As proposed in~\cite{Fleischhauer2020,Huang2015}, we use a combination of one Gamma kernel and one Gaussian kernel. Accordingly, a (discrete) pulse $x(k,\mathbf{\theta})$ is defined by 
\begin{equation}
    x\left(k,\mathbf{\theta} \right) = g_{\rm Gamma}\left(k,\mathbf{\theta}_{\rm Gamma}\right)+g_{\rm Gaussian}\left(k,\mathbf{\theta}_{\rm Gaussian}\right)
\end{equation}
where 
\begin{equation}
    g_{\rm Gamma}\left(k,\mathbf{\theta}_{\rm Gamma}\right)=am^{-mB} k^{mB} \exp{\left( \left(m-k\right)B \right)}
\end{equation}
with
\begin{equation}    
    B = \frac{1}{2\sigma}\left(m + \sqrt{m^2 + 4\sigma^2} \right)
\end{equation}
and
\begin{equation}
    g_{\rm Gaussian}\left(k,\mathbf{\theta}_{\rm Gaussian}\right) = a\exp\left( \frac{-\left(k-m\right)}{2\sigma}^2\right).
\end{equation}
The pulse is as a function of sample number $k$ with  $t =\frac{k}{f_{\rm s}}$ at sampling rate $f_{\rm s}$ and a parameter vector $\mathbf{\theta}$. Both kernels are parameterized by their amplitude $a$, position $m$ (mode) as well as standard deviation $\sigma$. The mode $m$ refers to the most common value in a probability distribution, i.e. the time of the maximum of the kernel. For a Gaussian kernel, the mode equals the mean. The standard deviation controls the kernel width. Note that the Gamma distribution originally is defined by parameter $\alpha$ and $\beta$ but can be expressed as a function of mode and standard deviation~\cite{Fleischhauer2020}. $\mathbf{\theta}$ is thus a three-parameter vector, which controls the actual pulse shape. To account for intersubject variability, we consider the different pulse classes, i.e. pulse shapes, according to Dawber et al.~\cite{Dawber1973}. The Dawber classes define four characteristic pulse wave shapes that reflect the continuum of possible pulse wave shapes. Each class is modelled with prototype values for the parameter vectors of either kernel (see Figure~\ref{fig:CLEAN_PPG} and Table~\ref{tab:PARAMETERS}). 

\begin{table}
    \centering
    \caption{Initial values of the parameter vectors $\mathbf{\theta}$ for the Gamma and Gaussian kernel to create the four prototype pulses according to Dawber's classes.}
    \footnotesize
    \begin{tabular}{c|c|c|c|c|c}
\textbf{kernel}	&	\textbf{parameter}	&	\textbf{class 1}	&	\textbf{class 2}	&	\textbf{class 3}	&	\textbf{class 4}	\\ \hline 
Gamma kernel	&	amplitude ($a$) / a.u.	&	0.9648	&	0.9623	&	0.9670	&	0.5384	\\
	&	mode ($m$) / s	&	0.1646	&	0.1836	&	0.2106	&	0.2162	\\
	&	standard deviation ($\sigma$) / s	&	0.0712	&	0.0839	&	0.1083	&	0.0924	\\
\hline
Gaussian kernel	&	amplitude ($a$) / a.u.	&	0.5466	&	0.4162	&	0.2563	&	0.5384	\\
	&	mode ($m$) / s	&	0.4278	&	0.4186	&	0.4290	&	0.3130	\\
	&	standard deviation ($\sigma$) / s	&	0.0924	&	0.0819	&	0.0672	&	0.1321	\\
\hline
    \end{tabular}
    \label{tab:PARAMETERS}
\end{table}

\subsubsection{Random pulse shape variations}
\label{sec:random_pulse_variations}
\begin{figure}
    \centering
    \includegraphics[width=15cm]{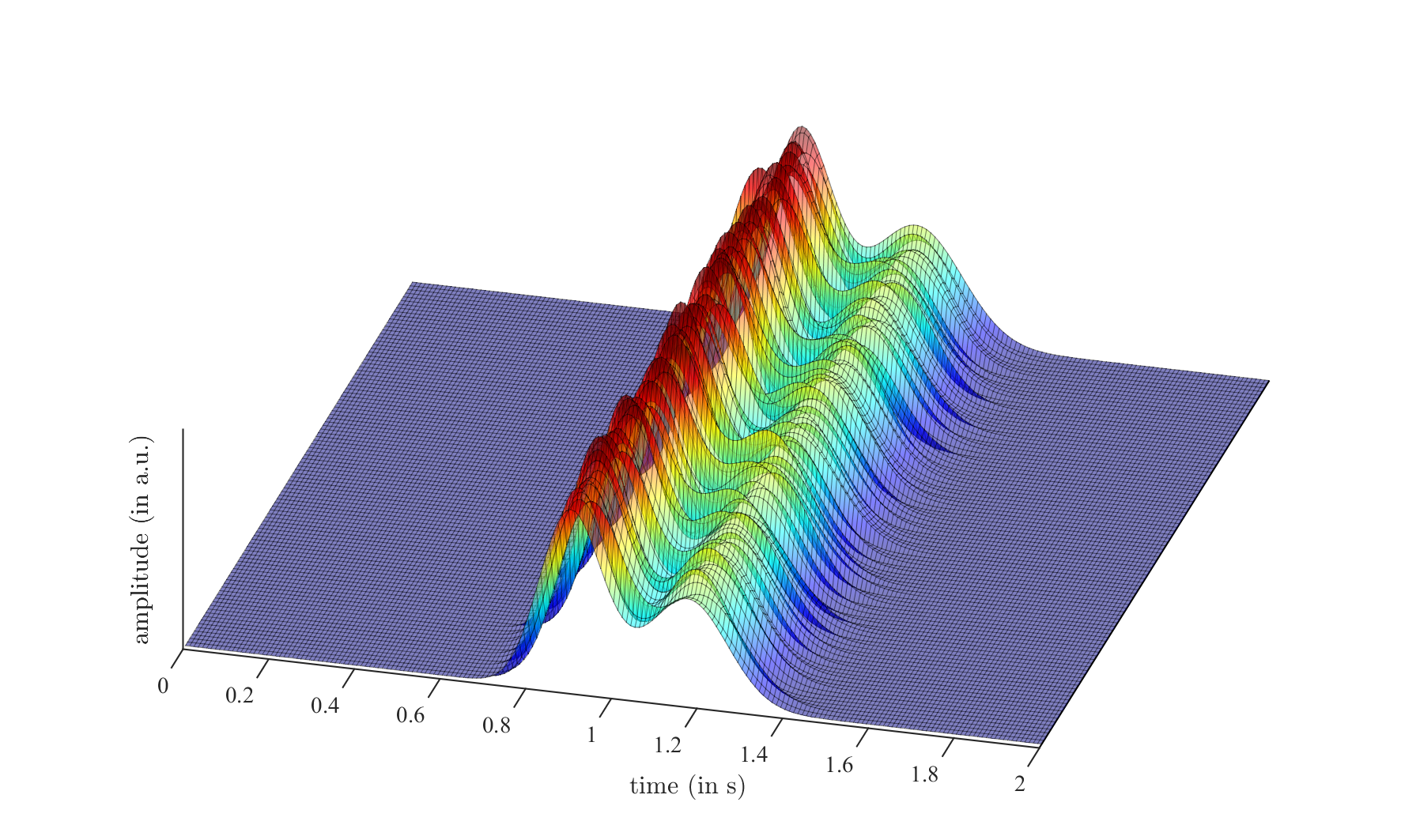}
    \caption{Illustration of random pulse variations in 100 prototype pulses before the addition of noise. Here, a Dawber class 1 beat is randomly varied. Distribution parameters were varied with standard deviation of 5\% (see text for details).}
    \label{fig:RANDOM_BEAT_VARIATIONS}
\end{figure}

Generating pulse prototypes by recomposition of kernels allows us to generate pulse shape variations by introducing fluctuations to the kernel parameters. Pulse shape variations can reflect both, physiological factors and intermittent disturbances including subtle movements and superposition of time-dependent effects of blood volume filling and ballistocardiography, which are common to IPPG~\cite{Moco2016a,Trumpp2017a}. To introduce pulse shape variations, we independently varied amplitude $a$ and width parameter $\sigma$ of used kernels by a percentage of its original values presented in Table~\ref{tab:PARAMETERS}. The percentage is randomly drawn from a normal distribution with the mean equal to zero and a predefined standard deviation of $\SI{5}{\%}$. Kernels’ positions $m$ were also varied. However, here the variation does not apply to the original positions as this would introduce larger variations to later occurring kernels. Instead, we varied the kernel position by a predefined percentage of the minimum distance to the neighbouring kernel(s)\footnote{Note that the exact same procedure could be applied to the use of multiple kernels.}. Figure~\ref{fig:RANDOM_BEAT_VARIATIONS} illustrates the effect of random variations at the \SI{5}{\%} level.

\subsubsection{Noise}
In order to investigate the impact of signal quality, we add pink noise to the prototype pulses to yield a predefined signal-to-noise ratio (SNR). We define the dimensionless SNR as the ratio of the symbol energy $E_{\rm S}$ to the noise power spectral density $N_0$,
\begin{equation}
    SNR =\frac{E_{\rm S}}{N_0}.
\end{equation}
The symbol energy $E_{\rm S}$ is the energy of the signal $x\left(k\right)$ divided by the number of samples $N_{\rm S}$ of the signal,
\begin{equation}
    E_{\rm S}=\frac{1}{N_{\rm S}}\sum_{k=1} ^{N_{\rm s}} x\left( k \right)^2.
\end{equation}

\begin{figure}
    \centering
    \includegraphics[trim=0.75cm 0cm 0.75cm 0cm, clip, width=\textwidth]{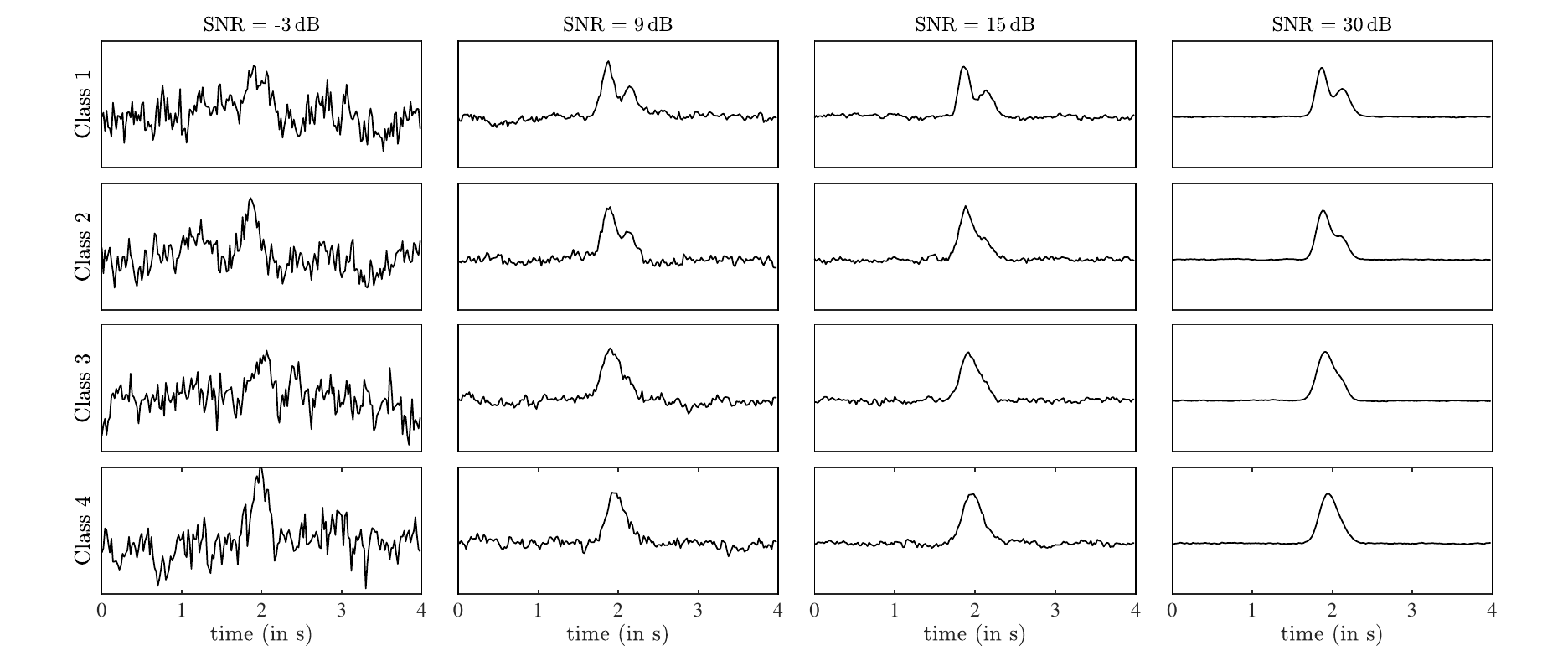}
    \caption{Visual impression of beats at variable SNR for all Dawber classes. As one can see, for an SNR of -3 dB, it is hard to distinguish the pulses from the background noise, whereas almost no background noise is visible for an SNR of 30 dB.}
    \label{fig:NOISE}
\end{figure}

We generate pink noise $n(k)$ by Fourier transforming a signal of standard normally distributed random numbers of the same length as $x(k)$ and manipulate the spectrum in such a way that the amplitude is proportional to $\frac{1}{\sqrt{f}}$. By applying an inverse Fourier transformation to the manipulated signal, ensuring unity standard deviation and a zero mean value and multiplying with $N_0$, we obtain $n(k)$ with the desired SNR (see \url{https://github.com/cortex-lab/MATLAB-tools/blob/master/pinknoise.m}). The noisy signal $y(k)$ is defined as the addition of the signal and the noise $y(k) = x(k) + n(k)$. Figure~\ref{fig:NOISE} gives a visual impression of four pulses from different Dawber classes at four used levels of noise. The shown signals cover a reasonable range of signal qualities, particularly considering low quality PPGI recordings or PPG signals from wearables.

\subsection{Monte Carlo Simulation}

To analyze the influence of sampling rate, SNR, and shape variations on BBI estimation error, we conduct a Monte Carlo Simulation. Algorithm~\ref{alg:MCS} provides the complete structure of the simulation. The text below details the procedure and used symbols. 

In each\revision{ realization of the random process }$i$, the simulation estimates the time delay of two pulses at a certain sampling rate and noise level and compares the estimate to the known true delay. Thereto, we first generate two prototype pulses $x_{\rm PPG,1}\left( k \right)$ and $ x_{\rm PPG,2}\left( k \right)$ according to Section~\ref{sec:prototype_generation}. The sampling rate $f^{\uparrow}_{\rm s}$ is set to $\SI{10000}{Hz}$ to emulate quasi-continuous signals. Assessing the PPG pulses at the exact same locations during downsampling (see below) might introduce unwanted (and unrealistic) systematic effects. We therefore introduce a random time delay to the pulses by convolving them with two impulse signals $\delta_1\left(k\right)$ and $\delta_2\left(k\right)$ of duration $t_0 = \SI{4}{s}$. For $\delta\left(k\right)$ holds
\begin{equation}
    \delta\left(k\right)=\begin{cases} 1 \quad \text{if} \quad k=k_0\\ 0 \quad \text{otherwise.}
    \end{cases} \quad k=\left\{ 0,1, \ldots, \SI{4}{s}\cdot f^{\uparrow}_{\rm s}\right\}
\end{equation}
In $\delta_1$, $k_0$ is set to equal the time instant $\frac{t_0}{2} = \SI{2}{s}$, i.e. $k_0=\frac{t_0}{2}\cdot f^{\uparrow}_{\rm s}$. In $\delta_2$, the impulse has a random delay $\Delta$, i.e. $k_0 = \frac{t_0}{2}\cdot f^{\uparrow}_{\rm s} + \Delta \cdot f^{\uparrow}_{\rm s}$ where $\Delta$ is normally distributed with $\mu_{\Delta}=0$ and $\sigma_{\Delta}=\SI{100}{ms}$. Such impulse signals are convolved with the prototype PPG pulses $x_{\rm PPG,1}\left( k \right)$ and $ x_{\rm PPG,2}\left( k \right)$  to yield the time delayed prototype pulses $x_1\left( k \right)$ and $x_2\left( k \right)$ according to
\begin{equation}
x_1\left( k \right) = \delta_1\left( k \right) \ast x_{\rm PPG,1}\left( k \right) \quad \text{and} \quad x_2\left( k \right) = \delta_2 \left( k \right) \ast x_{\rm PPG,2} \left( k \right).
\end{equation}
The signals are then truncated to be of the same length as $\delta_{1,2}$, i.e. \SI{4}{s}. Note that the random delay has only relevance to avoid downsampling to exact same sample points; otherwise we could work without delay, i.e. $\Delta=0$, and derive the error from delay estimates which typically will not be 0 due to the added noise. 


We consider two scenarios for prototype generation: in the first set of experiments (scenario 1), we do not introduce beat-to-beat shape variations to the pulse prototypes, i.e. both prototype pulses are identical $x_{\rm PPG,1} = x_{\rm PPG,2}$ with parameters as given in Table~\ref{tab:PARAMETERS}. In the second set of experiments (scenario 2), we introduce beat-to-beat shape variations according to the aforementioned procedure using a \SI{5}{\%} variation of kernel parameters.

Next, the two quasi-continuous time-delayed prototypes are downsampled to $x^{\downarrow}_1$ and $x^{\downarrow}_2$ with a reduced sampling frequency $f^{\downarrow}_{\rm s}$ using the MATLAB function $\texttt{resample}$. $\texttt{resample}$ applies a straight-forward FIR anti aliasing lowpass filter before performing the actual resampling. After downsampling, random noise is added to $x^{\downarrow}_{1}$ and $ x^{\downarrow}_{2}$ to obtain $\tilde{x}^{\downarrow}_{1}$ and $\tilde{x}^{\downarrow}_{2}$ having specific SNR. Finally, these signals are resampled to the BBI estimation frequency $f_{\rm s_{est}} = \SI{1000}{\hertz}$ to obtain $\hat{x}_{1}$ and $\hat{x}_{2}$ using $\texttt{resample}$ in MATLAB. The last step takes into account the positive effect upsampling described in the literature.

To estimate the delay $\Delta_{\rm est}$, the MATLAB function $\texttt{finddelay}$ is used. $\texttt{finddelay}$ calculates the cross-correlation between the two signals up to a lag of $\pm n_{\rm max}$ and then determines the shift between two signals by finding the lag-value that corresponds to the maximum cross-correlation. Here, $n_{\rm max}$ was set to correspond to one second to reduce computational time.\revision{ One second maximum displacement was chosen as the simulated beats’ effective length (i.e. samples that have a relevant difference from zero) is \SI{0.75}{s}. The expected value of shift is zero with a Gaussian distribution with $\sigma$=\SI{100}{ms}. A displacement $> \SI{1}{s}=10\sigma$ between beats would mean in any case that there is almost no overlap between samples that are different from zero between both beats. Displacements estimated in such a range can, even in real measurements, be easily identified as outliers from the mean pulse rate or mean pulse transit time (depending on the application). }In scenario 2, an additional estimation step is introduced. Owing to the shape variations, the cross-correlation of pulses would not likely yield the correct ${\Delta}$ even without noise, i.e. $\Delta_{\rm est} \neq {\Delta}$. In other words, shape variations introduce an additional error in the BBI estimate. To determine the effect of the shape variations, we estimated the delay $\hat{\Delta}$ between the two pulses using the quasi-continuous, noise-free signal (representing the delay which would be expected to be found without noise and thus the best possible result). This allows us to consider the impact of noise and shape variations separately. If both pulses are the same, $\Delta$ equals $\hat{\Delta}$.

At the end of each iteration $i$, the absolute error is stored to $AE(i) = \left|\Delta_{\rm est}(i)-\Delta(i)\right|$. In scenario 2, the compensated error is saved as well, $AE_{\rm C}(i) = \left|\Delta_{\rm est} - \hat{\Delta}(i)\right|$. This terminates each iteration, $i$ is increased and the next iteration is started with new prototypes, noise, and delay.

To terminate the Monte Carlo Simulation, the normalized standard error of the mean (NSEM) of the absolute estimation error accumulator was used. 
We define NSEM as 
\begin{equation}
    {\rm NSEM}(i) = \frac{{\rm sd}(AE)}{{\rm mean}(AE) \cdot \sqrt{i}}
\end{equation} 
with ``$\rm sd$'' being the standard deviation and ``$\rm mean$'' the mean of the accumulator. Note that NSEM deviates from the definition of the ``standard error of the mean'' by dividing by the sample mean to allow its application for different ranges of AE with one threshold value. Here, the simulation is terminated if NSEM falls below \SI{1}{\%}.

To analyze a wide range of parameters, the reduced sampling frequency $f^{\downarrow}_{\rm s}$ as well as the SNR were varied on a logarithmic scale. For $f^{\downarrow}_{\rm s}$, the values were \SI{5}{\hertz}, \SI{6}{\hertz}, \SI{8}{\hertz}, \SI{11}{\hertz}, \SI{14}{\hertz}, \SI{18}{\hertz}, \SI{23}{\hertz}, \SI{30}{\hertz}, \SI{39}{\hertz}, and \SI{50}{\hertz}. The SNR was swept from \SI{-3}{dB} to \SI{30}{dB} in \SI{3}{dB} increments.

\begin{algorithm}
	\caption{\label{alg:MCS}Basic structure of the used Monte Carlo Simulation. $\mathbf{\theta}$ defines the pulse class according to Table~\ref{tab:PARAMETERS}. If no random pulse shape variations are used, $\mathbf{\theta}_1=\mathbf{\theta}_2=\mathbf{\theta}$ and $AE_{\rm C}$ becomes obsolete as $AE = AE_{\rm C}$. Otherwise, $\mathbf{\theta}_1$ and $\mathbf{\theta}_2$ are random variations of $\mathbf{\theta}$ according to Section~\ref{sec:random_pulse_variations}. See text for further symbols' definitions and explanations.}  
	\textbf{Input:} $\mathbf{\theta}$ \Comment{PPG prototype parameters} \\
	\hspace*{\inputIntent} $\mathcal{SNR} \text{ (in dB)} =\left\{ -3,0,\ldots,30\right\}$ \Comment{assessed SNR}\\
	\hspace*{\inputIntent} $\mathcal{F}^{\downarrow}_{\rm s} \text{ (in Hz)} =\left\{5,6,8,11,14,18,23,30,39,50\right\}$ \Comment{assessed sampling rates}\\
    \textbf{Output: $AE, AE_{\rm C}$}  \Comment{absolute errors}
	\begin{algorithmic}[1]
			\For {$SNR \in \mathcal{SNR}$} \Comment{iteration over SNR}
			    \For {$f^{\downarrow}_{\rm s}  \in \mathcal{F}^{\downarrow}_{\rm s}$} \Comment{iteration over$f^{\downarrow}_{\rm s}$}
			        \State initialize empty arrays of $AE$ and $AE_{\rm C}$
			        \For {$i=1,\ldots,\infty$}\Comment{iteration over pairs of prototype pulses}
                        \State draw random delay $\Delta$
                        \State generate $\mathbf{\theta}_1$ and $\mathbf{\theta}_2$ from $\mathbf{\theta}$
                        \State generate signal $x_1$ from $\mathbf{\theta}_1$ at $t_0$
                        \State generate signal $x_2$ from $\mathbf{\theta}_2$ at $t_0 +  \Delta$
                        \State estimate $\hat{\Delta}$ from $x_{1}$ and $x_{2}$ using $\texttt{finddelay}$
                        \State downsample $x_{1}$ and $x_{2}$ at $f^{\downarrow}_{\rm s}$ to obtain $x^{\downarrow}_{1}$ and $x^{\downarrow}_{2}$
                        \State add noise at $SNR$ to $x^{\downarrow}_{1}$ and $ x^{\downarrow}_{2}$ to obtain $\tilde{x}^{\downarrow}_{1}$ and $\tilde{x}^{\downarrow}_{2}$
                        \State upsample $\tilde{x}^{\downarrow}_{1}$ and $ \tilde{x}^{\downarrow}_{2}$ at $f_{\rm s_{est}} = \SI{1000}{\hertz}$ to obtain $\hat{x}_{1}$ and $\hat{x}_{2}$
                        \State estimate $\Delta_{\rm est}$ from $\hat{x}_{1}$ and $\hat{x}_{2}$ using $\texttt{finddelay}$
                        \State append estimation errors $AE(i)$ and $AE_{\rm C}(i)$
                        \If{$i>2$ \textbf{and} $NSEM(i)<0.01$}
                            \State \textbf{break}
                        \EndIf
                    \EndFor
                    \State save $AE$ and $AE_{\rm C}$ for $SNR$ and $f^{\downarrow}_{\rm s}$
			    \EndFor
			\EndFor
	\end{algorithmic} 
\end{algorithm}

\subsection{Error estimation}
In order to quantify the resulting error per SNR and sampling rate, we use the root-mean-square error (RMSE) between the estimated BBI and the expected BBI. The RMSE is calculated by
\begin{equation}
    \text{RMSE BBI} = \sqrt{ \sum_{i=1} ^N\left( \Delta_{\rm est}\left( i \right) - \Delta \left( i \right)  \right)^2} =\sqrt{\sum_{i=1} ^N\left( AE\left(i\right) \right) ^2}
\end{equation}
or by
\begin{equation}
    \text{RMSE BBI} = \sqrt{\sum_{i=1} ^N\left( \Delta_{\rm est}\left( i \right)  - \hat{\Delta}\left( i \right)  \right)^2}= \sqrt{\sum_{i=1} ^N\left( AE_{\rm C}(i) \right)^2} 
\end{equation}
if we consider the effect of shape variations on the expected delay separately. $N$ is the number of iterations for respective sampling rate and SNR.

\section{Results}
\begin{figure}
    \centering
    \includegraphics[trim=2cm 0cm 1.5cm 0cm, clip, width=\textwidth]{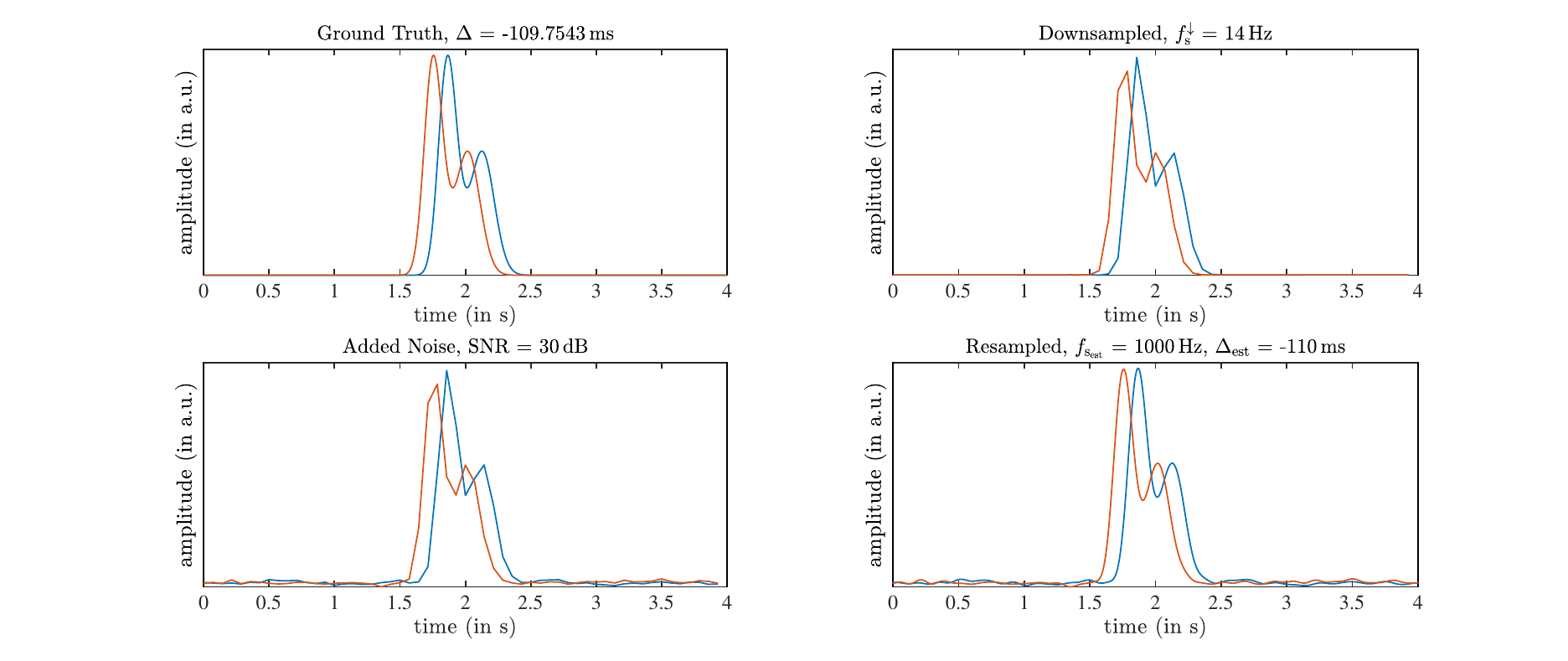}
    \caption{Illustration of a single iteration of the simulation without beat-to-beat shape variations. Top left: two time delayed pulse prototypes (equal shape); top right: downsampled version of the clean pulses; bottom left: downsampled pulses with added noise;  bottom right: resampled signals}
    \label{fig:EXAMPLE NORMAL}
\end{figure}

\begin{figure}
    \centering
    \includegraphics[trim=2cm 0cm 1.5cm 0cm, clip, width=\textwidth]{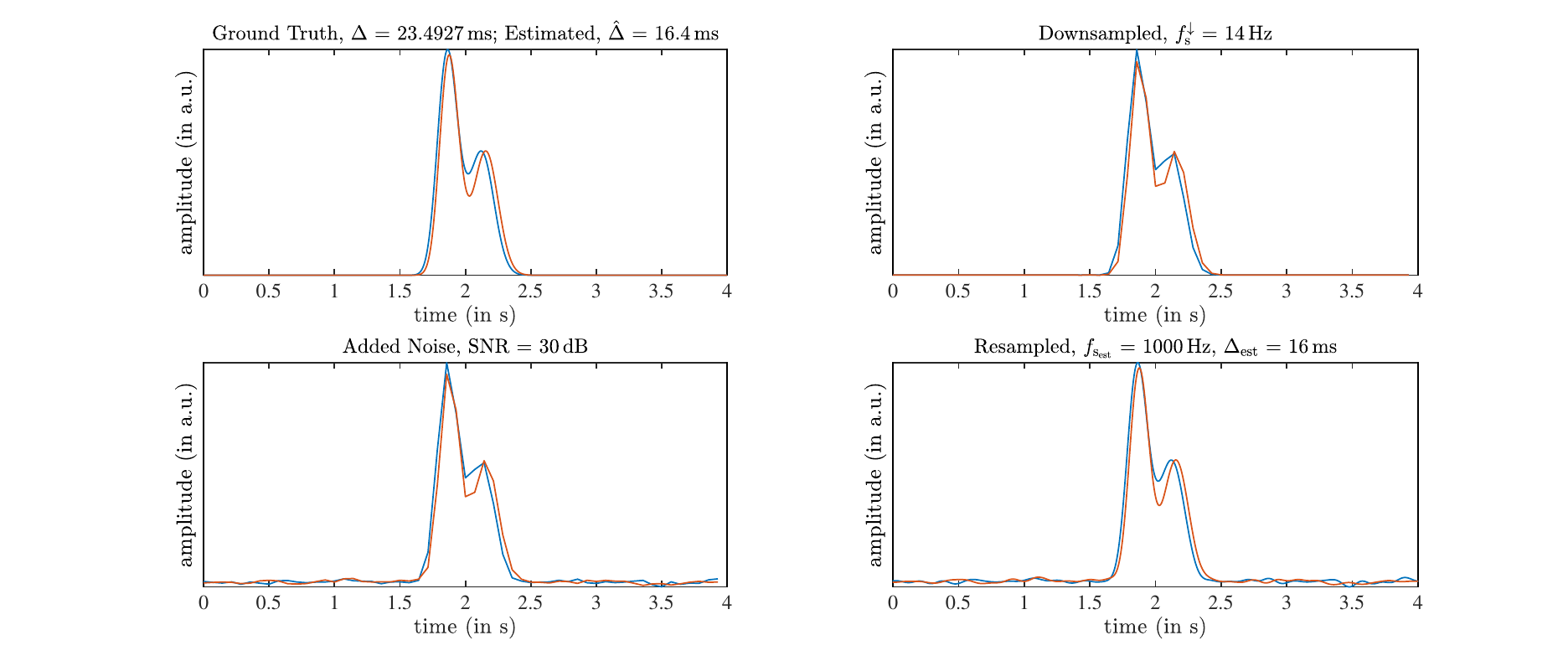}
    \caption{Illustration of a single iteration of the simulation with beat-to-beat shape variations. Left, upper row: two time delayed pulse prototypes (unequal shape); right, upper row: downsampled version of the clean pulses; left, lower row: downsampled pulses with added noise;  right, lower row: resampled signals}
    \label{fig:EXAMPLE RAND}
\end{figure}

\begin{figure}
    \centering
    \includegraphics[trim=2cm 0cm 1.5cm 0cm, clip, width=\textwidth]{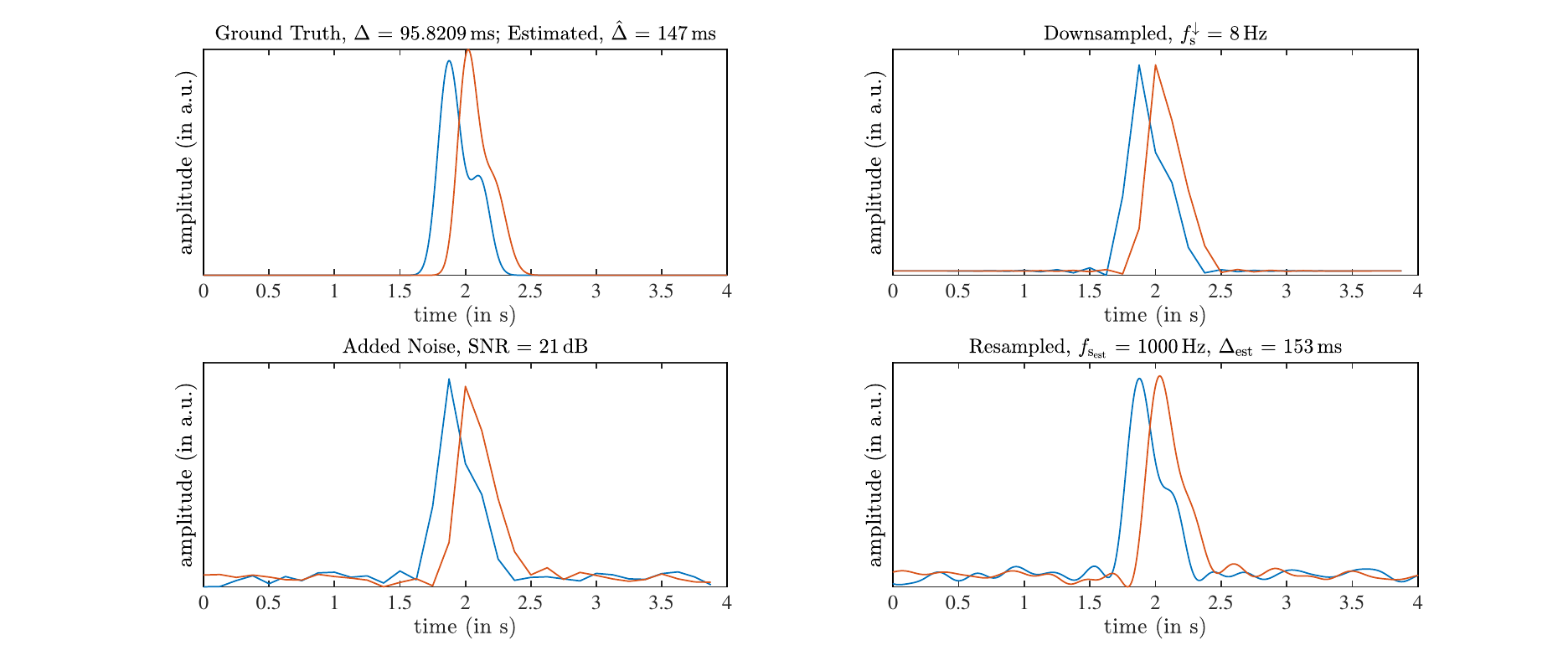}
    \caption{Illustration of a single iteration of the simulation with beat-to-beat shape variations with lower SNR and sampling rate compated to Figure~\ref{fig:EXAMPLE RAND}. Left, upper row: two time delayed pulse prototypes (unequal shape); right, upper row: downsampled version of the clean pulses; left, lower row: downsampled pulses with added noise;  right, lower row: resampled signals}
    \label{fig:EXAMPLE_RAND_LOW}
\end{figure}

Figures~\ref{fig:EXAMPLE NORMAL} and~\ref{fig:EXAMPLE RAND} illustrate one iteration of the Monte Carlo simulation. In both figures, class 1 beats are simulated and processed with a reduced sampling rate of $f^{\downarrow}_{\rm s} = \SI{14}{Hz}$ at an SNR of \SI{30}{dB}. In Figure~\ref{fig:EXAMPLE NORMAL},  both prototype pulses are the same (scenario 1), whereas \SI{5}{\%} shape variation in both beats is introduced in Figure~\ref{fig:EXAMPLE RAND} (scenario 2). Note that in both cases, the pulse shape can be recovered nicely by upsampling the downsampled signal. Further note that the subsequent delay estimation produces an error of less than \SI{0.25}{ms} in Figure~\ref{fig:EXAMPLE NORMAL} whereas the error exceeds \SI{4}{ms} in Figure~\ref{fig:EXAMPLE RAND} if we compare $\Delta_{\rm est}$ with $\Delta$. If we, however, compare it to $\hat{\Delta}$, we see that the difference is again minimal (\SI{0.5}{ms}). The situation is different in Figure~\ref{fig:EXAMPLE_RAND_LOW}, where both SNR and $f^{\downarrow}_{\rm s}$ are lower, \SI{21}{dB} and \SI{8}{Hz} respectively. Note that the dicrotic notch in the resampled blue curve is far less pronounced (bottom right). Also, even if we compare the estimation $\Delta_{\rm est}$ to the corrected $\hat{\Delta}$, the error is \SI{6}{ms}.  

\begin{figure}
    \centering
    \includegraphics[trim=0.25cm 0cm 0.25cm 0cm, clip, width=0.45\textwidth]{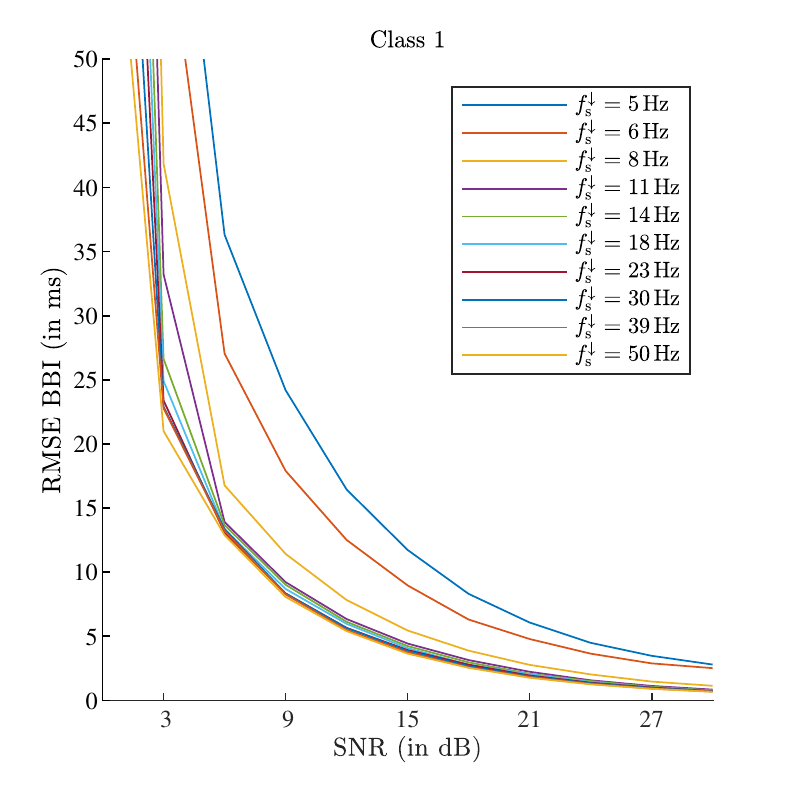}
    \includegraphics[trim=0.25cm 0cm 0.25cm 0cm, clip, width=0.45\textwidth]{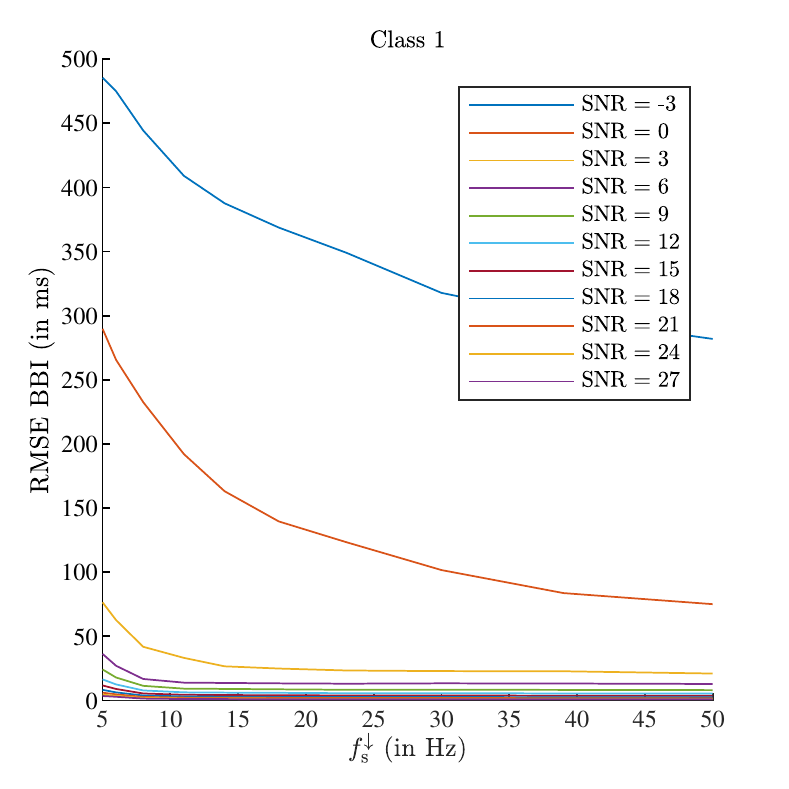}
    \includegraphics[trim=0.25cm 0cm 0.25cm 0cm, clip, width=0.99\textwidth]{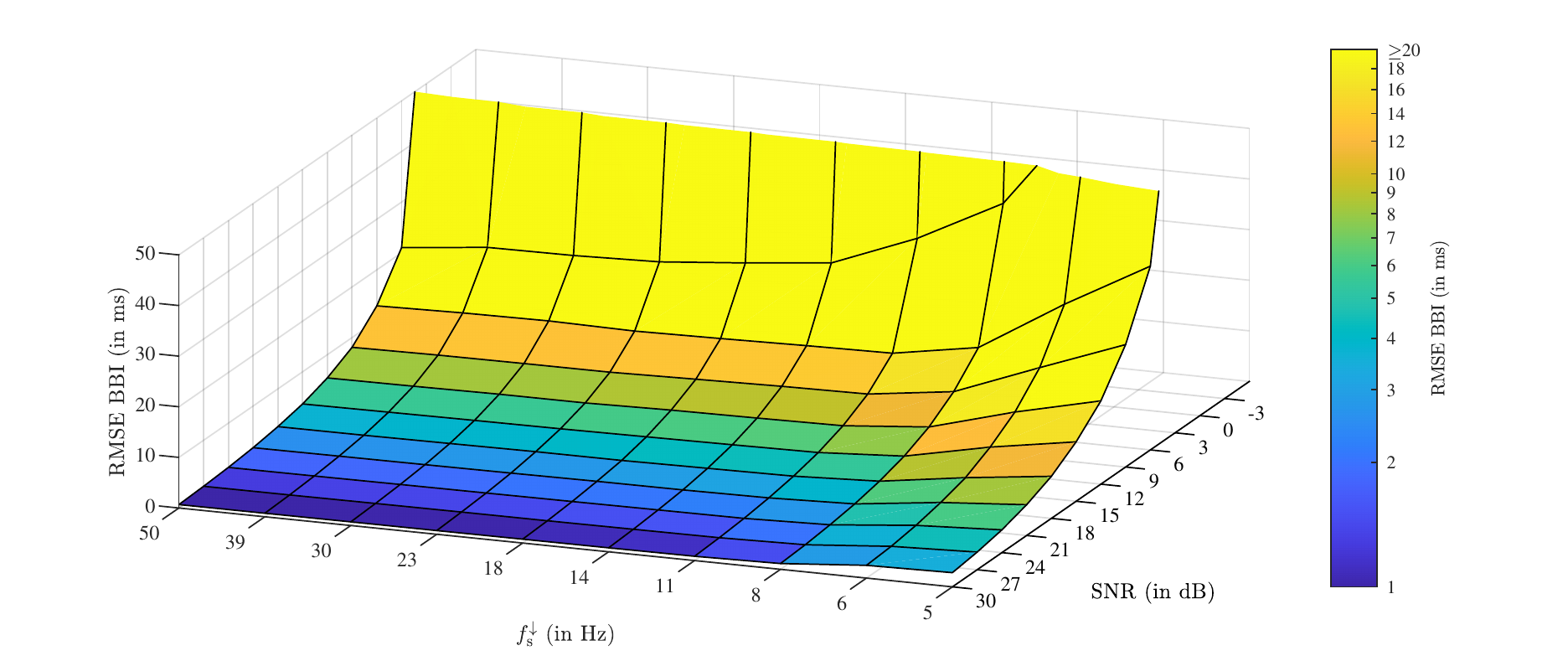}
    \caption{BBI root-mean-square error of the complete Monte Carlo simulation for class 1 beats without random variation. Note that the SNR has much larger impact than the sampling rate. \revision{The top left and the bottom panel are clipped at \SI{50}{ms} for better readability while the top right panel shows the full range of the data (all panels show the same information).}}
    \label{fig:CLASS 1 SWEEP}
\end{figure}

\revision{For NSEM to fall below \SI{1}{\%} and the simulation to terminate, a minimum / median / maximum number of runs of 4,979 / 5,701.5 / 47,770 were performed (all classes, no random variations). Most simulation runs were performed for the low-SNR scenarios with a median value of 18,655 runs for SNR $\leq$ \SI{3}{dB}. }In Figure~\ref{fig:CLASS 1 SWEEP}, the root-mean-square estimation error of the complete Monte Carlo simulation is shown for class 1 beats without random variation. With respect to the dependency to SNR, we can observe a strong decline of RMSE with an increase of the SNR. We also observe that the error is relatively high for $f^{\downarrow}_{\rm s}$ below \SI{8}{\hertz}. However, for $f^{\downarrow}_{\rm s}$ above \SI{8}{\hertz}, the difference between different sampling frequencies is overshadowed by the difference due to SNR.\revision{ Figure~\ref{fig:CLASS 1 SWEEP} also shows that the error is very large for SNRs below \SI{3}{dB}. }SNR generally dominates over $f^{\downarrow}_{\rm s}$.

\begin{figure}
    \centering
    \includegraphics[trim=1.5cm 0cm 1.5cm 0cm, clip, width=\textwidth]{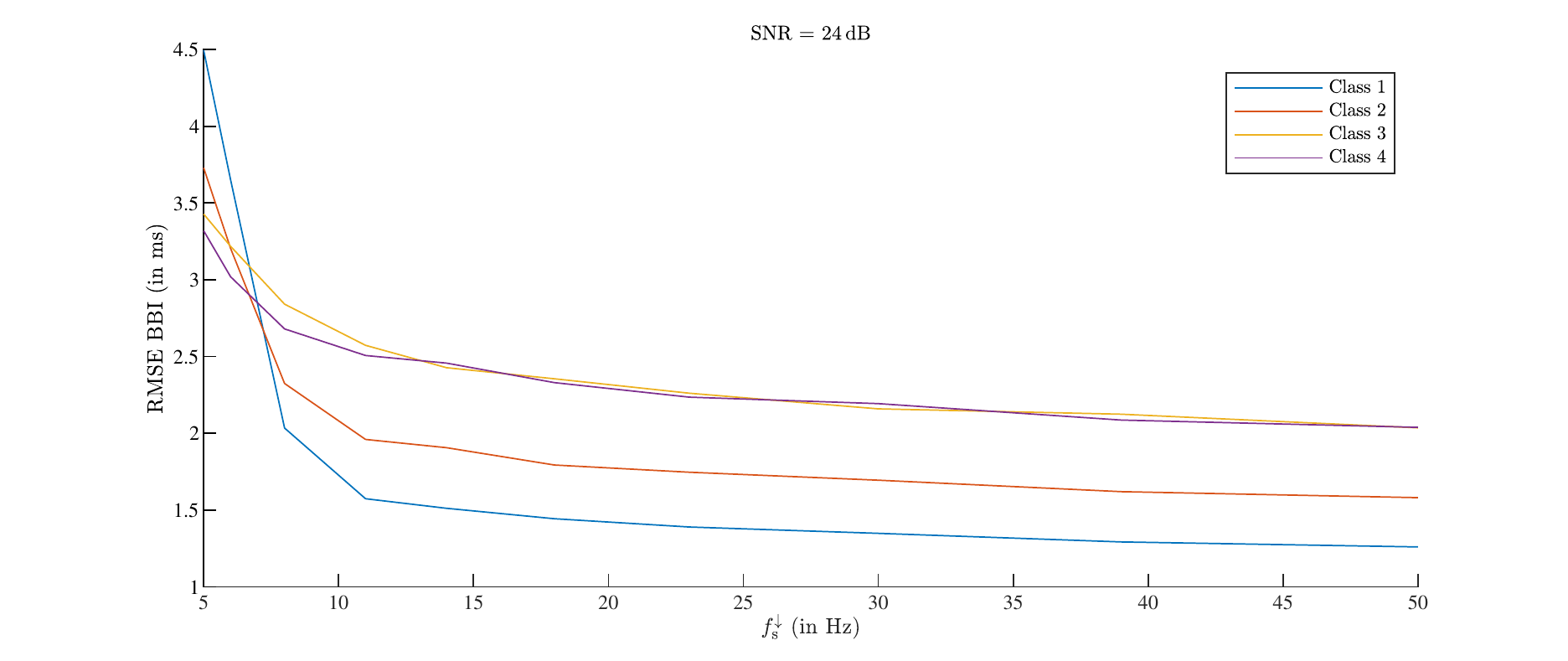}
    \caption{Root-mean-square error of the Monte Carlo simulation for all classes and a fixed SNR of \SI{24}{dB} (without random variation). Note that the error plateaus lowest for Dawber class 1 followed by class 2. The decrease in error is only marginal for an increase in sampling rate beyond \SI{15}{Hz} for all classes.}
    \label{fig:SNR FIX}
\end{figure}
\begin{figure}
    \centering
    \includegraphics[trim=1.5cm 0cm 1.5cm 0cm, clip, width=\textwidth]{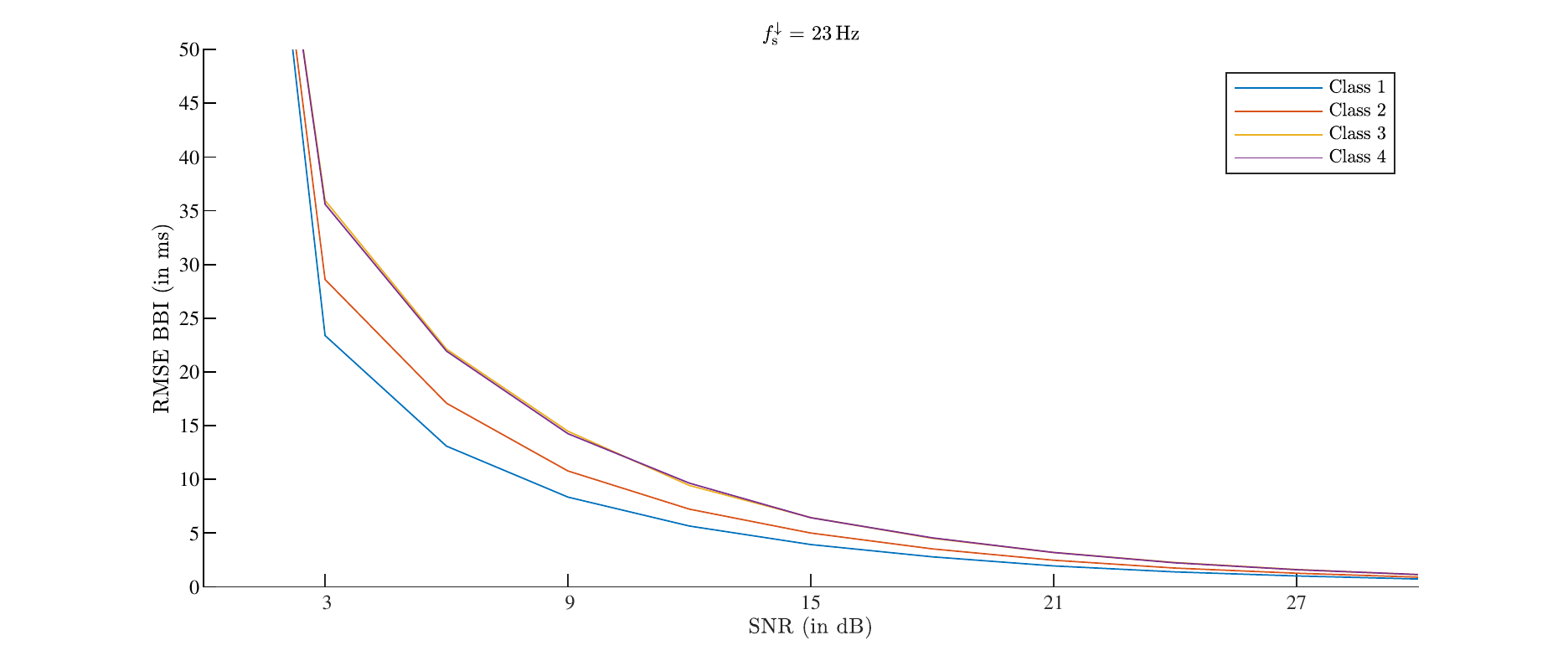}
    \caption{Root-mean-square error of the Monte Carlo simulation for all classes with a fixed $f_{\rm s}$ of \SI{23}{\hertz} (without random variation). For better readability, the y-axis is clipped at \SI{50}{ms}; the error peaks at approx. \SI{350}{ms} for class 1 and \SI{-3}{dB}.}
    \label{fig:FS FIX}
\end{figure}

In Figure~\ref{fig:SNR FIX}, for better readability, the SNR is fixed at \SI{24}{dB} while the sampling rate is swept from \SI{5}{\hertz} to \SI{50}{\hertz} as described above. In Figure~\ref{fig:FS FIX}, the sampling rate is fixed at $f^{\downarrow}_{\rm s} = \SI{23}{\hertz}$ while the SNR is swept. In addition to the information presented in Figure \ref{fig:CLASS 1 SWEEP}, the information for all Dawber classes, i.e. beat shapes, is added.

Figure \ref{fig:SNR FIX} shows that low sampling frequencies ($f^{\downarrow}_{\rm s} < \SI{10}{\hertz}$) increase the error dramatically. For very low sampling frequencies ($f^{\downarrow}_{\rm s} < \SI{7}{\hertz}$), we further see that the error is higher for lower pulse-classes (i.e. pulses with more high frequency components). For higher sampling rates ($f^{\downarrow}_{\rm s} > \SI{15}{\hertz}$), however, the estimation error is lower for lower pulse-classes. Moreover, we see that the decrease in error is only marginal for an increase in sampling rate beyond \SI{15}{Hz}. For example, the RMSE for class 1 at $f^{\downarrow}_{\rm s} = \SI{15}{\hertz}$ is approximately \SI{1.5}{ms}, while it is approximately \SI{1.3}{ms} for $f^{\downarrow}_{\rm s} = \SI{50}{\hertz}$.

\begin{figure}
    \centering
    \includegraphics[trim=1.5cm 0cm 1.5cm 0cm, clip, width=\textwidth]{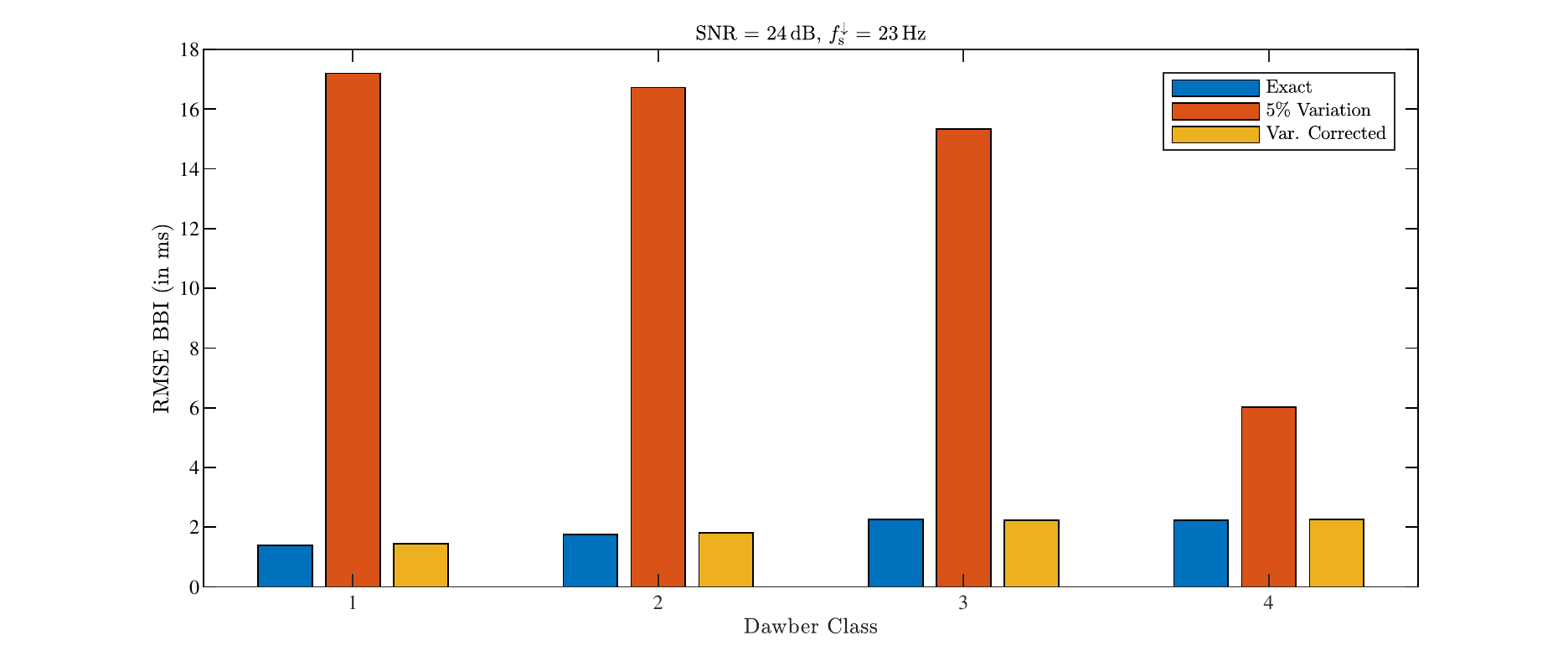}
    \caption{Comparison of exact simulations and random beat variations. With  \SI{5}{\%} variation in beat parameters, the error increases dramatically. If correction based on the noise-free, high-resolution signal is performed (``Var. Corrected''), the error returns to baseline.}
    \label{fig:NORM VS RAND}
\end{figure}

Finally, Figure~\ref{fig:NORM VS RAND} shows the RMSE for the four Dawber classes and in three cases where (a) both pulses have the same shape (``exact''), (b) the parameters of both pulses were varied by \SI{5}{\%} (``\SI{5}{\%} Variation'') and where $\Delta_{\rm est}$ is compared to the ground truth value $\Delta$, and finally (c) where $\Delta_{\rm est}$ is compared to $\hat{\Delta}$ (``Var. Corrected'').

First, it becomes obvious that the random variation of beat parameters in the range of \SI{5}{\%} increases the estimation error by a factor of almost 3 to more than 8, depending on the class. This error is notably smallest for Dawber class~4, which might be explained by the fact that most parts of the two kernels overlap and the pulse shape is thus less sensitive to variations in the parameters. At the same time, we can see that if we do not compare the estimated interval to the ground truth but to the value we achieve by performing cross correlation on the clean, continuous signal, we can see that the error returns to the level of two equal pulses. Thus, the large increase in error is independent of sampling rate and SNR.

\section{Discussion}
\subsection{Main findings and relation to other work}
First of all, the found errors should be put in relation to the expected temporal variation in PRV and pulse wave propagation to judge on the errors’ relevance. Regarding PRV, the root mean square of successive differences between normal heartbeats (RMSSD) is a widely used measure for heart rate variability and pulse rate variability, and is directly influenced by the beat-to-beat estimation uncertainty as captured by our RMSE. For\revision{ persons beyond the age of 55}, RMSSD is on average in the range of \SI{30}{ms} with a standard deviation in similar range~\cite{Tegegne2020} or even smaller~\cite{Umetani1998a}. E.g. infectious diseases usually slightly decreases RMSSD~\cite{Shaffer2017}. Found errors on the BBI estimation thus clearly are in a relevant range with respect to PRV estimation. Similarly, regarding applications focusing on analysis of pulse wave propagation, its variations \revdel{its variations }are in the order of milliseconds. E.g. (peripheral) PTT was $<\SI{60}{ms}$ in~\cite{Saiko2021, Fan2020} with a variation of approx.~\SI{20}{ms} upon blood pressure changes of approx.~\SI{20}{mmHg}~\cite{Fan2020}. PTT approaches to zero when close regions are considered~\cite{Saiko2021}. Accordingly, even regarding pulse wave propagation the found errors are relevant. 

The most important finding of our investigation is the distinguished impact of SNR. Previous works acknowledged the importance of SNR. E.g.~\cite{Pelaez-Coca2021} states ``\textit{[...] The accuracy of the PRV estimation is highly dependent on the possible signal interference or artefacts, and on the morphology of the PPG pulse, [...]}''. Bent and Dunn plan to include recordings under physical activity, and thus variable signal quality in future works~\cite{Bent2021}. Our own previous work has also shown that signal quality of PPG recorded from patients using wearable devices is quite volatile, and the quality of PRV derivation highly depends on both the specific PRV parameter and the individual patient~\cite{Hoog-Antink2021}. However, none of the comparable works did systematically consider SNR so far. Our investigations clearly prove a distinct impact of SNR. This finding is highly relevant in three regards: first, it can, at least partially, explain the differences existing in the literature on required sampling rates as recording conditions (hardware, protocols) and thus SNR varies. Second, it emphasizes the importance to consider SNR in future investigations on the sampling rate. Third, it underlines the importance of appropriate processing techniques to increase SNR in practical applications, particularly PPGI and wearable PPG. 

\begin{figure}
    \centering
    \begin{subfigure}[c]{7.5cm}
        \subcaption{Simulated pulse}
        \includegraphics[trim=0cm -.5cm 0cm 0.5cm, clip, width=\textwidth]{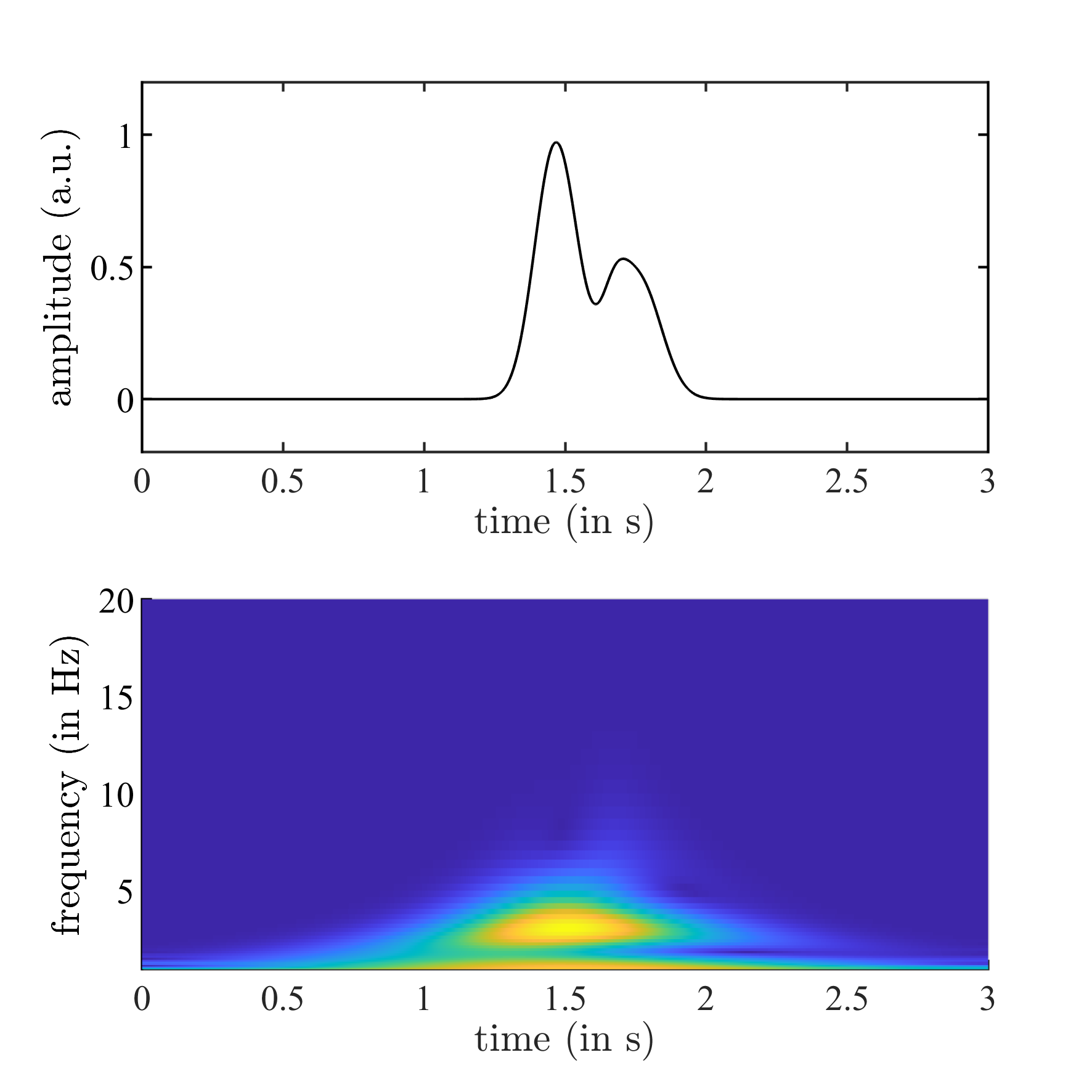}
    \end{subfigure}
    \begin{subfigure}[c]{7.5cm}
        \subcaption{Real pulse}    \includegraphics[trim=0cm 0.2cm 0cm 0.5cm, clip, width=\textwidth]{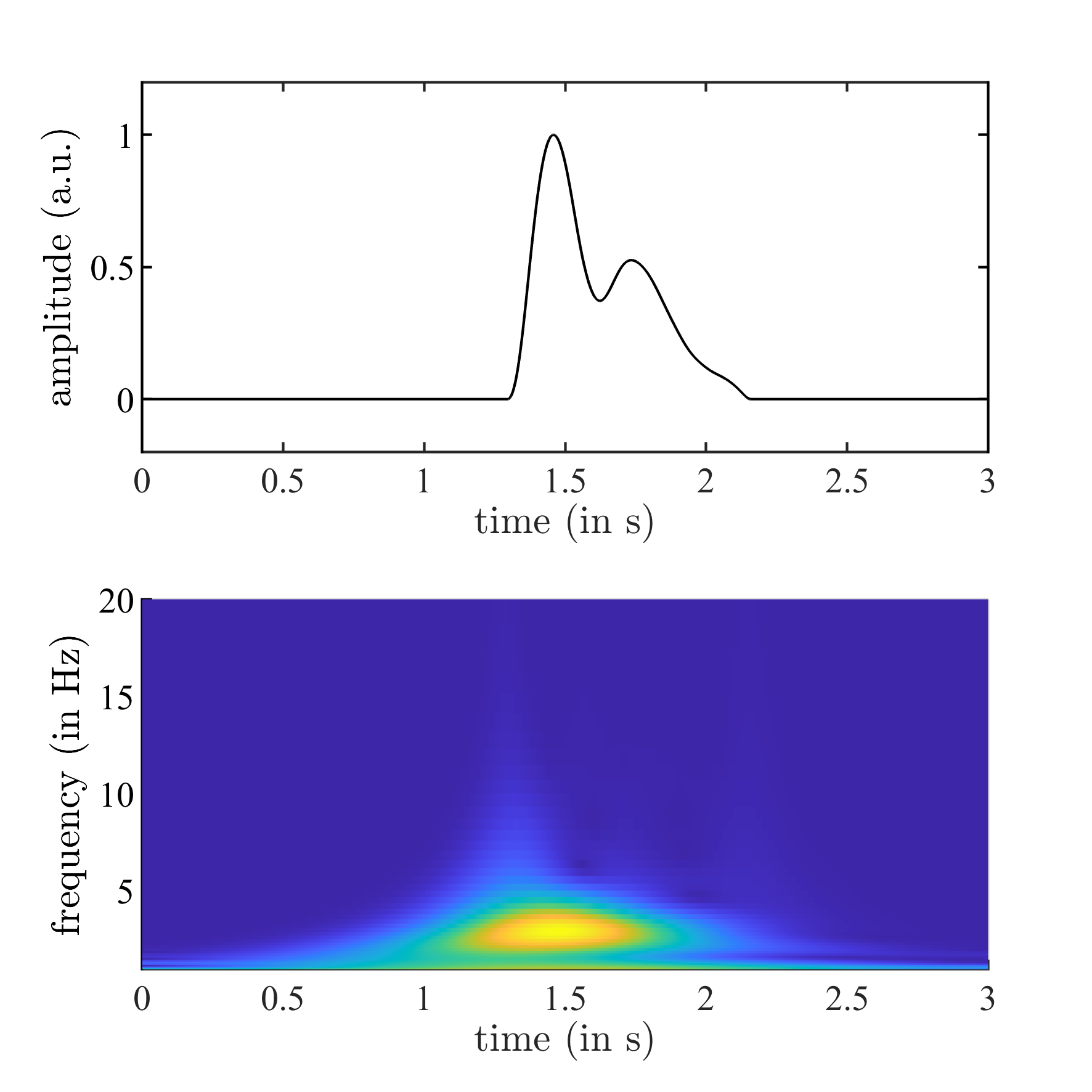}
    \end{subfigure}
    \caption{Illustration of a simulated and a real pulse (Dawber class 1). Upper panels show the time course. Lower panels show the time-frequency distribution (based on the analytic wavelet transform using the Morlet wavelet). The plots examplarily show the high degree of similarity.}
    \label{fig:FREQUENCY_CONTENT_ILLUSTRATION}
\end{figure}

With respect to the minimum sampling rate, our results are close to other investigations, which confirms the validity of our approach for signal modelling and error assessment. According to our investigation, the reduction in error flattens out at $f^{\downarrow}_{\rm s} > \SI{15}{Hz}$. Other investigations with real world data suggest slightly higher sampling frequencies (\SI{30}{Hz}~\cite{Fujita2019}, \SI{25}{Hz}~\cite{Choi2017a}, \SI{20}{Hz}~\cite{Beres2021}). A possible reason are the used pulse prototypes. It was shown in~\cite{Fleischhauer2020} and~\cite{Huang2015} that using one Gamma and one Gaussian kernel matches real pulse shapes very well. Figure~\ref{fig:FREQUENCY_CONTENT_ILLUSTRATION} illustrates the temporal and spectral similarity between a simulated beat and an exemplary real beat. Visually, the waveforms and the time-frequency distributions show a highly similar behaviour. The majority of the signal’s energy is concentrated well below \SI{7}{Hz}. However, in the time-frequency distribution of the real pulse signal, one can see a slightly higher frequency content in the upstroke at about \SI{1.3}{s} compared to the simulated pulse. Such subtle difference might explain the difference to the aforementioned works. Though there might be a small bias, even the visual impression on the error distribution over sampling frequency in~\cite{Choi2017} and~\cite{Bent2021}, which clearly becomes wider starting at \SI{20}{Hz}, closely relates to our findings. In any case, the subtle differences in frequency contents hardly explain why a sampling frequency of \SI{50}{Hz} to \SI{100}{Hz}  or \SI{64}{Hz} might be necessary as reported in~\cite{Pelaez-Coca2021} and~\cite{Bent2021}, respectively. Here, the definition of acceptable errors might play an essential role. This definition is difficult, e.g. owing to a dependence on contained BBI variability~\cite{Hejjel2017} (see Section~\ref{sec:limitations}).   

Further, we have shown that variations in beat morphology affect the estimation error. This applies to the beat class that affects the expected error but even more for beat-to-beat pulse shape variations. Such variations have a huge impact on the observed error. Our analysis demonstrates that this effect occurs independently of SNR and sampling rate. As with SNR, we are not aware of systematic investigation on this aspect so far. One could argue that the way we introduce variations and its strength (\SI{5}{\percent} variation of kernel parameters) are coarse approximation of reality. However, for example PPGI signals exhibit, owing to different measurement locations and locally effective distortions, a variability that far exceeds what we show in Figure~\ref{fig:RANDOM_BEAT_VARIATIONS} and have used in our experiments. As a consequence, we expect large errors in BBI estimations from PPGI and our findings need to be examined carefully in the future. 

Taken together, our findings are suited to impact the view on the minimal sampling rate and how to study it in the future. At the same time, the presented approach differs markedly from a previous simulation study on the topic~\cite{Beres2019} (the only one we are aware of) and provides much more detailed opportunities for signal modelling and error estimation under varying impact factors.

\subsection{Limitations and possible refinements}
\label{sec:limitations}
In our opinion, three aspects might be considered as limitations of our work.
First, our experiments are based on a pink noise model. In practice, other types of disturbances such as motion artifacts play an important role, too. Nevertheless, we believe that the general tendencies that we observe, particularly the SNR to be more relevant than the sampling rate, should hold under varied conditions. To support this hypothesis, we repeated our analysis using a white Gaussian noise model. The same tendencies were observed, with the error levels plateauing at lower levels, for example \SI{0.93}{ms} for white noise (not shown) vs.\SI{1.26}{ms} RMSE for pink noise (Figure~\ref{fig:SNR FIX}) at $f^{\downarrow}_{\rm s} =  \SI{50}{Hz}$ and $SNR = \SI{24}{dB}$. Also, for white noise, we observed that the error plateaued at higher sampling frequencies: for example, for the error to increase by approximately  \SI{10}{\percent} (\SI{1.26}{ms} to  \SI{1.39}{ms}), the sampling frequency could be lowered from $f^{\downarrow}_{\rm s} =  \SI{50}{Hz}$ to $f^{\downarrow}_{\rm s} =  \SI{23}{Hz}$ when using pink noise (Figure~\ref{fig:SNR FIX}). When white noise was used, however, a smaller decrease to $f^{\downarrow}_{\rm s} = \SI{39}{Hz}$ yielded the same relative increase in error (\SI{0.93}{ms} to  \SI{1.01}{ms}, not shown). These findings might also partly explain why previous works with real data came to slightly different findings with respect to minimal sampling rate, ranging from \SI{20}{Hz} to  \SI{30}{Hz}. We would speculate that more realistic noise models, e.g. introducing movement artifacts, might actually even surpass our finding of SNR's importance. This assumption is further supported by our experiments on beat-to-beat shape variations: such variations can be interpreted as a specific expression of noise and were shown to drastically increase the error as well. 

Second, we do not provide absolute statements on a minimum sampling rate as we do not answer the question of ``how good is good enough''. However, for some applications a BBI estimation error of \SI{5}{ms} might be tolerable while for others, \SI{1}{ms} might still be too high. Even the literature emphasizes different requirements on the sampling rate according to the population/pathology~\cite{Beres2019,Hejjel2017}, rendering any generalized statement on the minimum sampling rate questionable in our opinion. Against that background, we believe that we still demonstrated important findings, namely: an increase in sampling rate beyond a certain threshold, for example \SI{15}{Hz}, which might be costly in terms of battery life and/or storage space, will only result in limited improvements regarding BBI accuracy. At the same time, reducing the noise level, e.g. by proper illumination or sophisticated processing techniques, might be much more rewarding. We are aware that both parameters might be coupled as several samples might be temporally averaged to improve the SNR. This fact deserves further investigation: a hypothesis could be that storing of a de-noised, down-sampled signal is as beneficial as storing the original signal in practice.

\revision{Third, we do not employ any filtering of the noisy signal although it is possible that part of the noise energy involved in the SNR is easy to filter out. We acknowledge that filtering would be part of every real-world interval estimation approach. However, as no universally accepted preprocessing pipeline exists, we refrain from further preprocessing and suggest to interpret the SNR values as ``SNR after preprocessing'' as even after processing, some noise will always be left in the signal. Along the same lines}, our study relies on BBI estimation via cross-correlation. Other methods to estimate the  BBI - common choices would be the detection of local extrema or maximum slopes - might yield different results. However, we believe that the correlation method is an appropriate choice due to multiple reasons. First, correlation-based methods have been used quite successfully in the past to estimate BBI of different cardiac signals including PPG signals~\cite{Hoog-Antink2018}, which renders our specific choice reasonable. Second, other methods to estimate BBI that invoke multiple points (by curve fitting or temporal averaging) have been shown to be beneficial compared to using single points as local extrema or maximum slopes~\cite{Pelaez-Coca2021,Wanhua2021}. A correlation based approach is similar in the sense that multiple points determine the delay estimation rendering it reasonable again. Third, a widely used group of methods to extract spatio-temporal information from PPGI bases on the PPG signals' correlation to harmonic waves~\cite{Kamshilin2011}. Such close relation to the used correlation approach renders its usage again reasonable. These facts, combined with our intuition that BBI estimation using single points will be even more  susceptible to noise, makes us optimistic that our choice is reasonable and the general tendency will hold for other methods to estimate BBI as well.


\section{Conclusions}
Our investigations prove the SNR to deserve more attention regarding accurate BBI estimation in PPG signals. Considering our results, SNR is even more important than sampling rate. We consider this finding as highly relevant as, on the one hand, low SNR is common in current PPG applications such as PPGI and wearable PPG (e.g. optical heart rate monitors, smartwatches). On the other hand, typical procedures to investigate the minimum sampling rate do not consider the SNR, which might explain different findings to some extent and cause misleading conclusions. Therefore, we recommend a joint consideration of sampling rate and SNR in future works. In order to support respective approaches and own experiments in this regard, we released our data and scripts to the public domain (see \url{https://github.com/KISMED-TUDa/PPG_Sim_SNR_Fs}). Our work will also serve as basis for our own studies with real PPGI data and data from wearables. As an attempt to investigate the impact of SNR using real data, future works might include a signal quality estimator as proposed in~\cite{Li2012c}.

\section{Acknowledgements}
SZ and VF thank the Deutsche Forschungsgemeinschaft (DFG, German Research Foundation) (project 401786308) for funding this work.


\bibliography{library}


\end{document}